\documentclass{article}
\usepackage[left=1in, right=1in, top=1in, bottom=1in]{geometry}
\usepackage{graphicx}
\usepackage[small,bf,up]{caption}
\begin{document}
\title{Low Power Resonant Optical Excitation of an Optomechanical Cavity}
\author{Yiyang Gong$^{*}$, Armand Rundquist, Arka Majumdar, and Jelena Vu\v{c}kovi\'{c} \\
	\small\textit{Department of Electrical Engineering, Stanford University, Stanford, CA 94305} \\
	\small\textit{*email:yiyangg@stanford.edu}}
%\twocolumn[
\begin{@twocolumnfalse}
\maketitle
\begin{abstract}
We demonstrate the actuation of a double beam opto-mechanical cavity with a sinusoidally varying optical input power. We observe the driven mechanical motion with only 200 nW coupled to the optical cavity mode. We also investigate the pump power dependence of the radio-frequency response for both the driving power and the probe power. Finally, we investigate the dependence of the amplitude of the mechanical motion on mechanical cavity quality factor.
\end{abstract}
\end{@twocolumnfalse}
%]

%\doublespacing

Optomechanics, the study of the interaction between light and mechanical motion, has recently captured the imagination of photonics researchers \cite{Optomech_review,Kippenberg_sciencereview}. For example, researchers have probed radio frequency (RF) mechanical motion of nanometer sized objects \cite{Painter_Singlebeam,Painter_doublebeam}. In addition, proposals for using the optical gradient force to induce mechanical motion \cite{Povinelli_waveguide,Povinelli_ring, Tang_mechproposal} with \cite{Kippenberg_toroid,Painter_Singlebeam,Painter_forces,Notomi_doubleslab,Lipson_doublediskmech} and without \cite{Tang_mech1,Tang_mech,Roels_mech} the use of an optical cavity have been experimentally demonstrated. In fact, at very high optical and mechanical confinement \cite{Kippenberg_resolvedsideband,Painter_Singlebeam,Kippenberg_mechreview}, the amplitude of a mechanical mode can be greatly increased. In addition, at high input powers, regenerative mechanical oscillations occur, where the linewidth of the mechanical mode greatly decreases, while the amplitude of the mechanical oscillation greatly increases.

The experiments above have been done with continuous-wave (CW) excitation of an optical cavity or modulated excitation of a waveguide. However, the CW excitation mechanism requires the mechanical motion to induce an out-of-phase modulation of the laser input, as only those forces in quadrature with the mechanical motion perform mechanical work on the structure. Such effects are generally small, as the thermal motion of the structure only weakly perturbs the optical transmission properties of a waveguide or cavity. In the CW case, the amount of work done on the mechanical cavity is proportional to $\kappa^{-2}$, where $\kappa$ is the optical field decay rate in the cavity. However, an alternative to increase the transduction between optical power and mechanical motion is to use modulated pumping \cite{Tang_mech1,Tang_mech}. Such a scheme can do work that is proportional to $\kappa^{-1}$, and greatly reduce the amount of power needed to excite the mechanical mode (see Appendix). In this work, we demonstrate the use of optical pump modulation in conjunction with an optical cavity to reduce the amount of power needed to actuate the mechanical mode. Because of the optical confinement and recirculation of photons, we hope to obtain large mechanical oscillations without regenerative feedback.

In particular, we choose to work with the double beam one-dimensional photonic crystal (PC) cavity configuration in silicon. Due to its high optical quality factor ($Q > 10^4$), which enhances the circulating optical power inside the cavity, and low mode volume ($\sim (\lambda/n)^3$), which also enhances the local field potential, the PC cavity can greatly enhance the optical gradient force. The optomechanical coupling rate is defined as:
\begin{equation}
g_{\mbox{OM}} = \frac{d\omega}{dx},
\end{equation}
where $\omega$ is the optical cavity frequency and $x$ is the mechanical displacement of the cavity. By using cavities where the $E$-field is increased near material boundaries (such as in a slotted design \cite{Painter_doublebeam, Painter_2DPCslot}), the frequency perturbation with mechanical motion and the optomechanical coupling can both be tailored. 

\begin{figure}[ht!]
\centering
\includegraphics[width=2.7in]{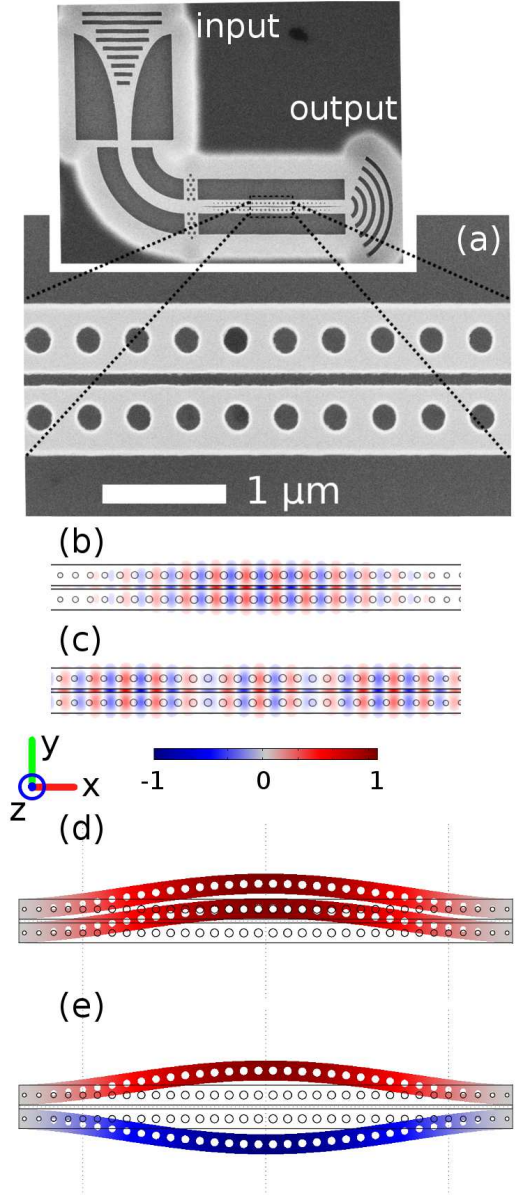}
\caption{(a) Scanning electron microscope image of the fabricated cavity. The $E_{y}$ field of the (b) TE$_{1,+}$ and (c) TE$_{2,+}$ optical modes. (d) The first order common in-plane mechanical mode, and (e) the first order differential in-plane mechanical modes are plotted with the color map assigned to the in-plane ($y$) motion.}
\label{fig:modes}
\end{figure}

We fabricate devices on a silicon-on-insulator (SOI) wafer with a 150 nm thick layer of Si and a 1 $\mu$m thick oxide layer, such as the cavity shown in Fig. \ref{fig:modes}(a). The beam cavities have lengths of approximately 13 $\mu$m, single beam widths of 550 nm, and a middle slot width of 100 nm. We use the design of Ref. \cite{Quan_taperout}, where the hole lattice constant is kept constant at $a=400$ nm, and the radii of circular holes are reduced as the distance from the center of the cavity increases. The larger holes at the center of the cavity create an optical potential well that lies in the optical bandgap of the array of outer holes, and such a design allows robust and high-efficiency coupling to the cavity region via a coupling waveguide. The hole at the center of the cavity has radius $r = 0.28a$, and the total cavity length is 34 holes. The cavity is fabricated with electron-beam lithography, and the pattern is transferred into the silicon layer by a Cl$_{2}$:HBr plasma dry etch. The oxide sacrificial layer is then etched away using a buffered oxide etch (BOE) to obtain the free standing beams. In addition to the beam cavity, we also attach coupling waveguides on both sides of the cavity, and one of the waveguides is bent 90$^{\circ}$ to configure the device to be probed in a cross-polarization geometry [Fig. \ref{fig:modes}(a)]  \cite{Hatice_ccavs,Dirk_refl}.

% (Sibeam 6 7 1, set 4,3 cav 3,4, parameters ends up being a = 400nm, r/a = 0.28, s = 100 nm, w  = 550nm.)

We first simulate the beam cavities in the optical regime using the three dimensional finite-difference time-domain (3D-FDTD) method. Double beam cavities support bonded ($+$) and anti-bonded ($-$) optical super-modes, formed from the the transverse-electric (TE) modes of the individual beam cavities. In particular, the $E_{y}$ field is symmetric or anti-symmetric about the $xz$-plane going through the slot for the bonded and anti-bonded modes, respectively. We find that the first (TE$_{1,+}$) and second (TE$_{2,+}$) order bonded modes (see Fig \ref{fig:modes}(b)-(c)) have theoretical radiation-limited $Q$s of 30,000 and 1,500, respectively. We observe an enhanced electric field in the air slot region for the bonded modes because of the continuity conditions for the dominant $E_{y}$ field at the slot boundaries (i.e. continuity of the displacement vector $\epsilon \vec{E}$). Thus, we expect that the bonded optical modes have the highest optomechanical coupling to the in-plane mechanical modes, as the high electric field concentration in the middle of the cavity enhances the change in the optical cavity frequency with mechanical deformations. For this reason, we work with the first and second order bonded optical modes in our experiments.

\begin{figure}[ht]
\centering
\includegraphics[width=3.5in]{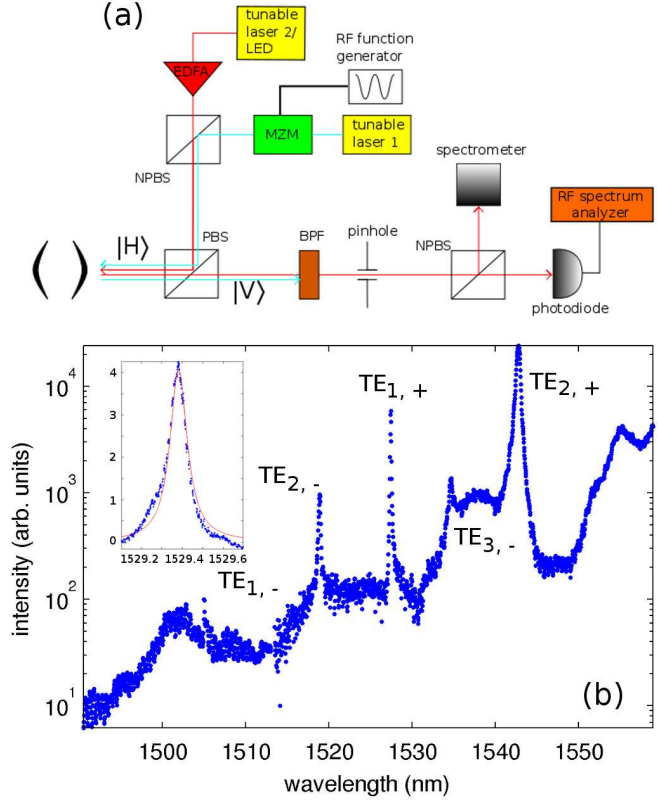}
\caption{(a) The optical setup used to probe the optomechanical cavity. (b) Spectrum of the cavity observed in transmission using a broadband LED. The first and second order bonded ($+$) and anti-bonded ($-$) modes are labeled. The inset shows a laser scan of the TE$_{1,+}$ cavity mode for excitation, with a fit to a Lorentzian lineshape having $Q\approx15,000$.}
\label{fig:spec}
\end{figure}

We experimentally analyze the optical properties of the cavity using the setup in Fig. \ref{fig:spec}(a). We pump the cavities with a broadband LED bank, which is coupled into a waveguide using a dielectric grating coupler. We align the cavity such that the input grating polarization is along $\left|H\right>$, while the output polarization is along $\left|V\right>$, to obtain the maximum signal to noise ratio. The transmission characteristics of the cavity are shown in Fig. \ref{fig:spec}(b), where we are able to observe the first two orders of the bonded and the anti-bonded modes. We are able to differentiate the bonded modes from the anti-bonded modes by moving the input beam on the grating coupler to change the input parity.  The first order modes have high $Q$-factors, and we use a tunable laser to fully characterize the cavity. The laser scan at low input powers (1 nW) shows a Lorentzian spectrum with $Q \approx 15,000$ for the bonded first order mode (TE$_{1,+}$) (inset Fig. \ref{fig:spec}(b)). In addition, we observe that the higher order bonded mode (TE$_{2,+}$) has $Q\approx 2,000$. Both $Q$ values are comparable to the FDTD simulated values.

We next use the COMSOL finite element solver to find the frequencies of the mechanical modes, using library parameters for silicon: Young's modulus of 131 GPa, Poisson's ratio of 0.27, and a density of 2.33 g/cm$^{3}$. As described above and in previous work \cite{Painter_doublebeam}, mechanical modes with in-plane (in this case, referring to the $xy$ plane) motion will have significant optomechanical coupling to the bonded modes. In particular, we find the first order common and differential modes for in-plane motion \cite{Painter_doublebeam}. The common and differential modes have the beams moving in phase and out of phase, primarily in the $y$-direction, and have displacement profiles shown in Fig. \ref{fig:modes}(d) and (e), respectively. By simulating the structure observed in the SEM image, we find that these two mechanical modes have mechanical frequencies of 25.72 MHz and 26.74 MHz. We find the optomechanical coupling strength similarly to previous work \cite{Painter_Singlebeam}, with the optomechanical coupling length defined as:
\begin{equation}
\frac{1}{L_{\mbox{OM}}} = \frac{1}{2}\frac{\int dA \left(\frac{dq}{d\alpha} \cdot \hat{n} \right) \left( \Delta\epsilon |E_{||}|^2-\Delta(\epsilon^{-1})|D_{\perp}|^2 \right)}  {\int dV \epsilon |E|^2}.
\end{equation}
Here, $q$ is the mechanical displacement, $\alpha$ is the parameterized displacement of the mechanical mode, $\hat{n}$ is the surface normal vector, $E_{||}$ is the electric field parallel to the surface, $D_{\perp}$ is displacement field normal to the surface, $\Delta\epsilon = \epsilon_{1}-\epsilon_{2}$, and $\Delta(\epsilon)^{-1} = \epsilon_{1}^{-1}-\epsilon_{2}^{-1}$, with $\epsilon_{1}$ being the dielectric constant of silicon, and $\epsilon_{2}$ the dielectric constant of the surrounding medium.  Because of the high $E$-field enhancement in the slot and the differential mechanical resonance having opposite parity to the $E_{y}$ field, we observe very strong optomechanical coupling lengths of $L_{\mbox{OM}} = $ 1.3 $\mu$m and 1.8 $\mu$m for the coupling between the differential mechanical mode and the TE$_{1,+}$ and TE$_{2,+}$ optical modes, respectively. On the other hand, the coupling between the TE$_{1,+}$ and TE$_{2,+}$ optical modes and the common mechanical mode was calculated to be far weaker ($L_{\mbox{OM}} >$ 40 $\mu$m), because this mechanical mode has the same parity as the optical field.

\begin{figure}[ht]
\centering
\includegraphics[width=3.5in]{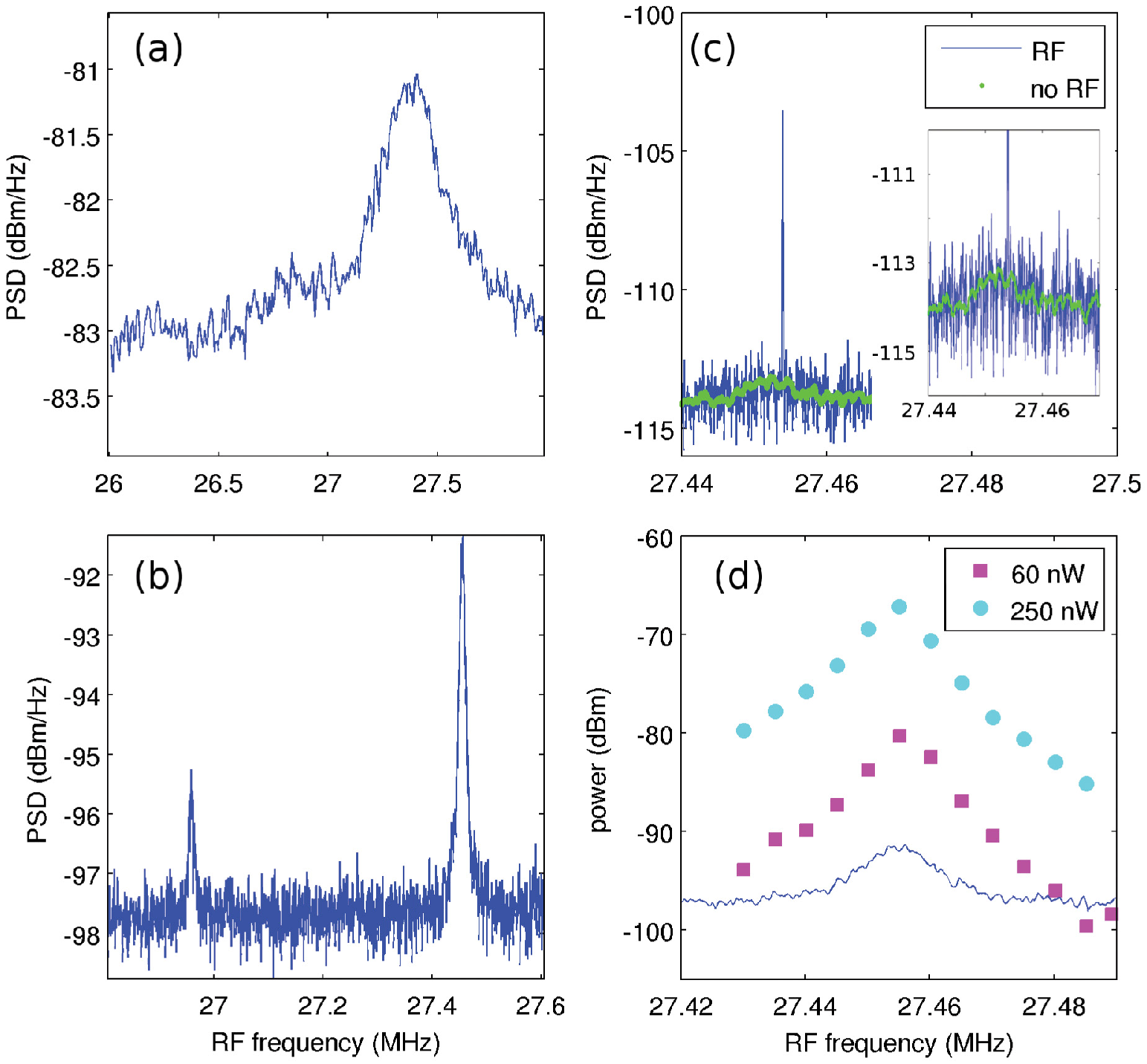}
\caption{The RF spectrum of the mechanical modes under study in (a) ambient atmosphere, and in (b) vacuum. (c) The time averaged spectrum of the differential mechanical mode from part (b) is shown (green points), observed as RF sidebands of the laser tuned to TE$_{2,+}$. The non-averaged RF spectrum showing the sharp RF response when a modulated laser on TE$_{1,+}$ is added is also plotted (blue line). The inset shows the same data zoomed in, to observe the thermal driven mechanical mode in the background. (d) The integrated power within the sharp RF response of the laser on TE$_{2,+}$ [from (c)] with different RF modulation frequencies of the laser on TE$_{1,+}$. The two dotted curves correspond to two different average input powers on the first order mode and fixed input power on the second order mode. A closer zoom of the mode shown in part (b) of the figure is shown as a reference at the bottom (blue).}
\label{fig:RFmodes}
\end{figure}

In order to first characterize the mechanical modes of the system, we pump the second order bonded mode with a red detuned probe laser, at the cavity half-max, with low pump power (300 $\mu$W before the objective) to observe the mechanical modes in air. The transmission signal is fiber-coupled and sent to a photodiode detector with a transimpedance gain of $2.5 \times 10^4$ V/A and a bandwidth of 125 MHz, and the electrical signal is then read by an RF spectrum analyzer. We estimate coupling efficiencies of 2\% to the TE$_{2,+}$ mode and 0.5\% to the TE$_{1,+}$ mode, assuming symmetric losses at the input and output gratings, and accounting for the transmission losses of the coupling waveguides using FDTD simulations. We observe the two mechanical modes in the RF spectrum, shown in Fig. \ref{fig:RFmodes}(a), which correspond well to the simulated in-plane mechanical mode frequencies, and slight discrepancies can be attributed to minor differences in the clamping conditions of the fabricated device. Because of the low optical $Q$ of the TE$_{2,+}$ mode and the low optical power buildup, we do not observe the giant optical spring effect seen in previous works \cite{Painter_doublebeam,Lipson_doublediskmech}, as the mechanical modes do not change frequency with increasing pump power. We also do not observe significant changes in the optical cavity wavelength with pump power, suggesting minimal heating.  Because of the mechanical damping of the ambient atmosphere, the mechanical $Q$-factors of these modes are limited to 50-100. When we test the same cavity in vacuum, we observe the two modes more clearly, as shown in Fig. \ref{fig:RFmodes}(b). In vacuum, the mechanical $Q$s are as high as 2,500, and are limited by the clamping geometry of our cavity. We choose to work with the higher frequency mode (the differential mode), as it is the in-plane mechanical mode with higher optomechanical coupling to the second order optical mode.

Next, we pump the TE$_{1,+}$ mode with a second (pump) laser tuned to the optical cavity resonance wavelength and sinusoidally modulated near the RF frequency of the mechanical mode, while keeping the first CW laser tuned to the half-maximum of TE$_{2,+}$. We observe the effect of the second, modulated laser on the RF modulation of the first laser. We scan through the first-order optical mode with various unmodulated powers, and observe that the first order cavity resonance is not significantly changed, suggesting that the injected power on the first order optical mode does not change the temperature of the beam, and thus does not modulate the beam transmission via the thermo-optic effect. Although both lasers pass through the cavity and are extracted with the same output grating coupler, the laser on TE$_{1,+}$ is blocked by a band-pass filter centered at 1550 nm with a full-width at half-max of 12 nm. The power of the laser on TE$_{1,+}$ is modulated by a Mach-Zender interferometer modulator with a bandwidth of 2.5 GHz and full modulation depth [Fig. \ref{fig:spec}(a)]. First, we fix the input power on the first-order optical mode at under 2 $\mu$W, and scan the modulation frequency through the mechanical resonance. When we tune the RF input frequency near the mechanical resonance frequency, we observe a narrow response in the RF spectrum (of the laser on TE$_{2,+}$) [Fig. \ref{fig:RFmodes}(c)]. In addition, as the RF input frequency is tuned around the mechanical resonance frequency, we observe that the integrated power within the narrow bandwidth response matches exactly that of the mechanical cavity resonance [Fig. \ref{fig:RFmodes}(d)], suggesting that the optical power in the first order mode is modulating the transmission properties of the second order mode through the mechanical resonance. In addition, we observe the Lorentzian mechanical mode with far better signal to noise, and can observe the tails of the mechanical mode even when detuned by more than three mechanical cavity linewidths. 

\begin{figure}[ht]
\centering
\includegraphics[width=3.5in]{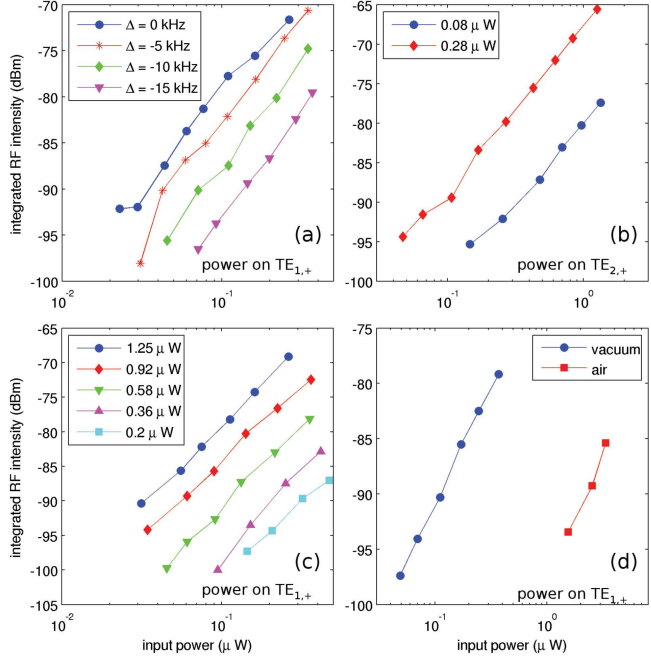}
\caption{(a) The integrated intensity in the RF response collected from TE$_{2,+}$ as a function of average input power on TE$_{1,+}$ for different detunings of the RF modulation frequency from the mechanical resonance at a fixed probe power (2 $\mu$W) on TE$_{2,+}$. (b) The integrated intensity in the RF response as a function of different probe powers on TE$_{2,+}$, at two different fixed average pump powers on TE$_{1,+}$. (c) The RF response as a function of input power on TE$_{1,+}$ with different probe pump powers on TE$_{2,+}$. (d) The integrated RF response as a function of average pump power on TE$_{1,+}$. The two curves correspond to the response at ambient atmosphere and in vacuum, both with the same probe intensity on the TE$_{2,+}$ mode (2 $\mu$W).}
\label{fig:RFseries}
\end{figure}

We also measure the RF response of the probe laser on TE$_{2,+}$ as we change the power of the modulated pump laser on TE$_{1,+}$. We first do so with the probe power for TE$_{2,+}$ fixed at 2 $\mu W$ coupled into the cavity, and observe the RF response with varying average power on TE$_{1,+}$ for different RF detunings from the mechanical resonance [plotted on a log-log scale in Fig \ref{fig:RFseries}(a)]. Similar to the data in Fig. \ref{fig:RFmodes}(d), we observe the RF response is decreased as the modulation frequency is detuned from the mechanical resonance. We observe that the relationship between the integrated power in the RF response and the input laser power on TE$_{1,+}$ is quadratic for all detunings. This is expected, as the RF spectrum analyzer measures the power of the voltage signal from the transimpedance amplifier of our detector, and that power has a quadratic relationship with the amplifier output voltage and thus a quadratic relationship with the output RF oscillation amplitude. This indicates a linear relationship between displacement and input pump power on the first order mode. 

We also measure the RF power spectrum from TE$_{2,+}$ when we fix the average laser power on TE$_{1,+}$, and increase the power of the pump on the second order mode, as shown in Fig \ref{fig:RFseries}(b). Again, we observe that the integrated RF response of the driven mechanical mode is quadratic with the input power, which is expected as the sideband amplitude is linearly related to the probe power. We also obtain the RF response as a function of the input power on TE$_{1,+}$ for various probe powers on TE$_{2,+}$, shown in Fig \ref{fig:RFseries}(c). The RF response is reduced for lower input powers, as the sideband powers are proportional to the input probe power. However, we are able to observe an RF response with only 100 nW coupled to the TE$_{1,+}$ mode to drive the mechanical oscillations, and only 200 nW coupled to the TE$_{2,+}$ mode to sense the mechanical motion.

Finally, we compare the efficiency of exciting the mechanical mode in vacuum and in ambient atmosphere. We fix the input power for the probe laser on the TE$_{2,+}$ mode in both air and vacuum to 2 $\mu$W, and obtain the same output coupled power into our photodetector. We obtain the power series from the same cavity under both conditions, which is shown in Fig. \ref{fig:RFseries}(d). As expected, the amplitude of the mechanical oscillation is significantly higher in vacuum than in ambient atmosphere, due to the higher mechanical $Q$. In fact, the experimentally measured factor of 20 between the power needed to generate the same RF response in air and vacuum matches well with the ratio of mechanical $Q$s for the two conditions (31).

In conclusion, we have demonstrated resonant actuation of a mechanical mode with optical gradient forces. The input power needed to observe driven motion of the mechanical cavity is greatly decreased in the presence of an optical cavity, and hundreds of nanowatts can drive the mechanical motion via a modulated laser coupled to a second cavity mode. This type of excitation can be used to probe various mechanical modes, as the RF response can be increased relative to the thermal-driven oscillations. Furthermore, optomechanical cavities can be used to mix RF signals, with the mechanical resonance enhancing the beat note of two RF signals. Similarly, the actuation of mechanical motion can also be used for a variety of applications, such as mechanical motors that do work on nanometer-sized objects. 

We acknowledge support from the Presidential Early Career Award for Science and Engineering (PECASE), administered by the Office of Naval Research (Dr. Chagaan Baatar). We also acknowledge support from the National Science Foundation graduate research fellowship (YG), and the Stanford Graduate Fellowship (AR, AM).

\section{Appendix: Theory of resonant excitation of mechanical mode with optical gradient force:}
We would like to solve for the mechanical amplitude as a function of the average input power of a modulated laser. We follow the derivation given in Ref. \cite{Painter_Singlebeam} and start with the cavity field equation:
\begin{equation}
\dot{c}(t) = -\left(\frac{\kappa}{2}+i\omega_{0}\right)c(t)+\frac{i\alpha(t)\omega_{0}}{L_{OM}}c(t)+\sqrt{\frac{\kappa_{e}}{2}}s(t)e^{-i\omega t}
\end{equation}
where $s(t)$ is the time-varying pump field, $\omega_{0}$ is the cavity frequency, $\kappa$ is the cavity field decay rate, $\kappa_{e}$ is the external coupling rate, $L_{OM}$ is the optomechanical coupling, $c(t)$ is the cavity field, and $\alpha(t)$ is the mechanical mode amplitude. In this case, we are inputing a laser at $\omega$ which is detuned from the optical cavity mode center frequency, and the input is modulated periodically with frequency $\Omega$, which is detuned from the mechanical mode center frequency $\Omega_{0}$.

We assume sinusoidal mechanical motion, such that the beam also moves with modulation frequency $\Omega$:
\begin{equation}
\alpha(t) = \alpha_{0}\sin(\Omega t)
\end{equation}
Note that $\Omega$ could be different from $\Omega_{0}$, but since we're driving the motion, we can assume the mechanical mode responds with the same frequency. Then the equation becomes:
\begin{equation}
\dot{c}(t) = -\left(\frac{\kappa}{2}+i\omega_{0}\right)c(t)+\frac{i\omega_{0}\alpha_{0}\sin(\Omega t)}{L_{OM}}c(t)+\sqrt{\frac{\kappa_{e}}{2}}s(t)e^{-i\omega t}
\end{equation}

The homogeneous solution is:
\begin{eqnarray}
c_h(t) = C_{0}/u = C_{0}\exp\left(-(\frac{\kappa}{2}+i\omega_{0})t - \frac{i\omega_{0}\alpha_{0}\cos(\Omega t)}{L_{OM}\Omega}\right) \nonumber \\
 = C_{0}\exp\left(-(\frac{\kappa}{2}+i\omega_{0})t\right) \sum\limits_{n}{(-i)^n J_{n}(\beta)e^{in\Omega t}}
\end{eqnarray}
with $\beta = \omega_{0}\alpha_{0}/L_{OM}\Omega$, and the inhomogeneous solution is:
\begin{equation}
c_{p}(t) = \int{u \sqrt{\frac{\kappa_{e}}{2}}s(t)e^{-i\omega t}}
=\int{e^{(\frac{\kappa}{2}+i\omega_{0})t} \sum\limits_{n}{i^n J_{n}(\beta)e^{in\Omega t}} \sqrt{\frac{\kappa_{e}}{2}}s(t)e^{-i\omega t}}
\end{equation}

Since our pump is modulated with frequency $\Omega$, we express $s(t)=\sum\limits_{k}{a_{k}}e^{ik\Omega t}$ as a Fourier Series, and find the full inhomogeneous solution:
\begin{eqnarray}
c_{p}(t) & = & \int{e^{(\frac{\kappa}{2}+i\omega_{0})t} \sum\limits_{n}{i^n J_{n}(\beta)e^{in\Omega t}} \sqrt{\frac{\kappa_{e}}{2}}\sum\limits_{k}{a_{k}}e^{ik\Omega t}e^{-i\omega t}}dt \nonumber \\
& = & \sum\limits_{n,k}{\frac{i^n J_{n}(\beta)a_{k}}{\frac{\kappa}{2}-i\Delta+i(n+k)\Omega} e^{(-i\Delta+i(n+k)\Omega-i\omega_{0})t - i\beta \cos(\Omega t)}}
\label{eqn:sumequation}
\end{eqnarray}
with $\Delta = \omega-\omega_{0}$, and neglecting the $\sqrt{\kappa_{e}/2}$ term as normalization:

Because the homogeneous solution levels out with rate $\kappa$ and this is fast, the particular solution is the steady state solution. The optical force is:
\begin{eqnarray}
\frac{|c_p(t)|^2}{L_{OM}} = \frac{1}{L_{OM}}\sum\limits_{n,k,m,l}{\frac{i^{n-m} J_{n}(\beta) J_{m}(\beta) a_{k}a^{*}_{l}}{(\frac{\kappa}{2}-i\Delta+i(n+k)\Omega)(\frac{\kappa}{2}+i\Delta-i(m+l)\Omega)} e^{i[(n+k)-(m+l)]\Omega t}}
\label{eqn:sinepump}
\end{eqnarray}

Taking only the zeroth order in $J_{0}(\beta)$, as $\beta \ll 1$ and $J_{1}(\beta) \approx \beta$:
\begin{eqnarray}
\frac{|c_p(t)|^2}{L_{OM}} = \frac{1}{L_{OM}}\sum\limits_{k,l}{\frac{J_{0}^2(\beta) a_{k} a^{*}_{l}}{(\frac{\kappa}{2}-i\Delta+ik\Omega)(\frac{\kappa}{2}+i\Delta-il\Omega)}
 e^{i(k-l)\Omega t}}
\end{eqnarray}

\subsection{Example 1: Cosine input}
Let's input $s(t) = s_{0}(1+\cos(\Omega t))/2$:
\begin{equation}
s(t) = s_{0}\left(\frac{1}{2}+\frac{1}{4}e^{i \Omega t}+\frac{1}{4}e^{-i \Omega t}\right)
\end{equation}
Thus we have $a_{0} = 1/2$, $a_{1}=a_{-1}=1/4$.

The normalization for the time dependent portion of the input is $A^2\int|s(t)|^2 = A^2\int(1+\cos(\Omega t))^2/4=A^2(3\pi)/(4\Omega)$. We want to keep the average power the same, so $A^2(3\pi)/(4\Omega)/T = 1 = A^2(3\pi)/(4\Omega)/(2\pi/\Omega)$, or $A = \sqrt{8/3}$.

The optical force, normalized to the average input power is then:
\begin{eqnarray}
\frac{F}{|s_{0}|^2 \kappa_{e} A^2} = \frac{J_{0}^2(\beta)}{L_{OM}}\sum\limits_{k,l}{\frac{a_{k} a^{*}_{l}}{(\frac{\kappa}{2}-i\Delta+ik\Omega)(\frac{\kappa}{2}+i\Delta-il\Omega)}
 e^{i(k-l)\Omega t}}
\end{eqnarray}

We will only consider the elements with frequency $\Omega$, in quadrature with the beam motion, as they will contribute to work getting done on the mechanical mode, so we isolate the $\cos(\Omega t)$ terms:
\begin{eqnarray}
\frac{F}{|s_{0}|^2 \kappa_{e} A^2} && = \frac{J_{0}^2(\beta)}{4 L_{OM}} \cos(\Omega t) \Bigg[ \frac{1}{\frac{\kappa^2}{4}+(\Delta-\Omega)^2} + \frac{1}{\frac{\kappa^2}{4}+(\Delta+\Omega)^2} \nonumber \\
&& - \frac{\Delta\Omega}{(\frac{\kappa^2}{4}+\Delta^2)(\frac{\kappa^2}{4}+(\Delta-\Omega)^2)}
+\frac{\Delta\Omega}{(\frac{\kappa^2}{4}+\Delta^2)(\frac{\kappa^2}{4}+(\Delta+\Omega)^2)} \Bigg] 
\end{eqnarray}

Note that this force is maximized near $\Delta = 0$ (as all four terms are near Lorentzian functions in terms of $\Delta$), and we consider the force amplitude (dropping the harmonic variation):
\begin{equation}
\frac{F}{|s_{0}|^2 \kappa_{e} A^2} = \frac{J_{0}^2(\beta)}{2 L_{OM}}\left[ \frac{1}{\frac{\kappa^2}{4}+\Omega^2}\right] \approx \frac{1}{2 L_{OM}}\left[ \frac{1}{\frac{\kappa^2}{4}+\Omega^2}\right]
\end{equation}
or with the normalization (such that input power is proportional to $|s|^2$):
\begin{equation}
\frac{F}{|s_{0}|^2 \kappa_{e}} = \frac{8}{3}\frac{1}{2 L_{OM}} \frac{1}{\frac{\kappa^2}{4}+\Omega^2}
\end{equation}

We note that in the case of pumping with a CW laser, the equivalent force is \cite{Painter_Singlebeam}:
\begin{equation}
\frac{F}{|s_{0}|^2 \kappa_{e}} = \frac{\beta}{2 L_{OM}}\left[ \frac{4\kappa\Delta\Omega^2}{(\frac{\kappa^2}{4}+\Delta^2)(\frac{\kappa^2}{4}+(\Delta-\Omega)^2)(\frac{\kappa^2}{4}+(\Delta+\Omega)^2)}\right]
\end{equation}

\begin{figure}[ht!]
\centering
\includegraphics[width=3.0in]{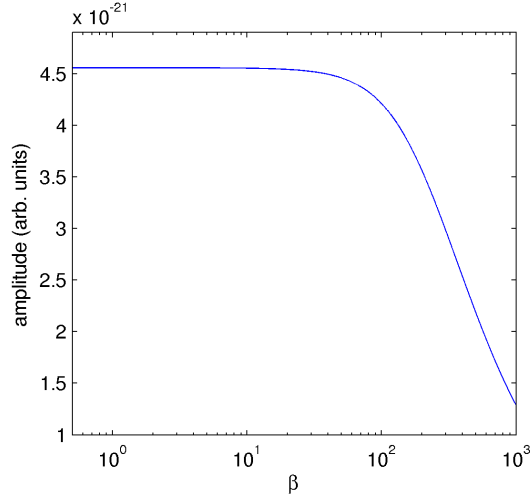}
\caption{The theoretical average force on the mechanical mode for a fixed average input power of the modulated input, as a function of $\beta$.}
\label{fig:highbeta}
\end{figure}

Thus, comparing some sort of AC pump scheme (assuming $\Delta = 0$, to maximize force) to the DC pumping (assuming $\Delta=\kappa/2$, where the force is approximately maximized), we see that the transferred power should be approximately $\kappa^2/(\beta\Omega^2)$ more efficient. In addition, if we assume that our in-coupling efficiency is sufficiently high, then we would have $\kappa_{e} \sim \kappa$. Using the above two equations, our optical force is $\propto \kappa^{-2}$ for the CW case, and $\propto \kappa^{-1}$ for the modulated laser case.

Note that our thermal amplitude is $\left<x^2\right> = k_{b}T/m_{\mbox{eff}}\Omega^2$, and $\left|\left<x\right>\right| \approx $ 10 pm in this case, which places us in the high $\beta$ regime (despite the sidebands being unresolved, we have $\beta = 45$). We can calculate the force as a function of $\beta$ as well, plotted in Fig. \ref{fig:highbeta}, using real parameters of $m_{eff}=2\times10^{-15}$ kg, $\Omega_{0} = 2\pi \cdot 22 \times 10^{6}$ Hz, $Q_{m}=70$, $\kappa_{e} = \kappa/2$, $L_{OM}=$2 $\mu$m, optical wavelength $\lambda = 1500$ nm, and optical $Q = 2\times10^4$. We plot the kernel of the force term using Eqn. \ref{eqn:sinepump} for different $\beta$, but fixing all other parameters, and plot the results in Fig. \ref{fig:highbeta}. We notice that for our parameters, the force on the beam is relatively unchanged even up to $\beta \approx 100$. Thus, we use sinusoidal pump to increase the force amplitude.

\subsection{Example 2: Square wave input}
We can also explore input powers that are periodic with frequency $\Omega$, but not sinuisoidal. One example is a square wave input that is that is zero for some amount of time, and a fixed amplitude $A=\sqrt{\frac{T}{2T_{1}}}$ (chosen to have fixed energy input) for time $2T_{1}$. The pulse train is input with period $T=2\pi/\Omega$.
\begin{eqnarray}
s(t) = \left\{ \begin{array}{rl}
 A &\mbox{ if $|t|<T_{1}$} \\
  0 &\mbox{ otherwise}
       \end{array} \right.
\end{eqnarray}

\begin{figure}[ht!]
\centering
\includegraphics[width=3.5in]{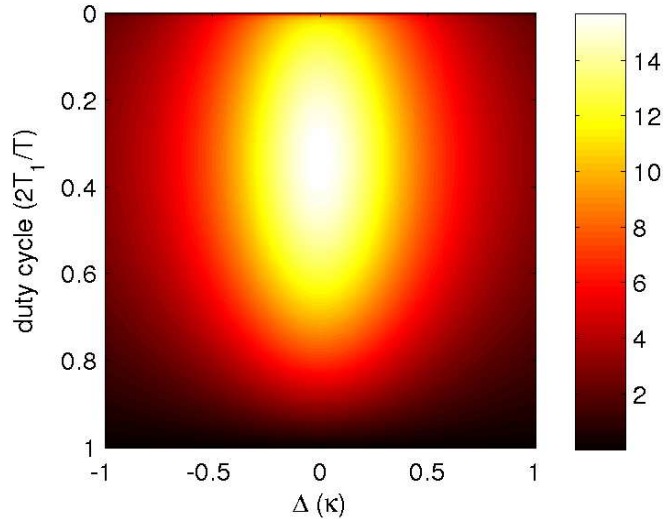}
\caption{The scaled force for fixed input energy, as a function of the duty cycle of the pump, and the optical detuning from the cavity.}
\label{fig:duty}
\end{figure}

The Fourier coefficients of this input are:
\begin{eqnarray}
a_{k} = \left\{ \begin{array}{rl}
 2AT_{1}/T &\mbox{ if $k = 0$} \\
  A \frac{\mbox{sin}(k\frac{2\pi}{T}T_{1})}{k\pi} &\mbox{ otherwise}
       \end{array} \right.
\end{eqnarray}

By evaluating the sum numerically, we obtain the force as a function of the duty cycle ($2T_{1}/T$) in Fig. \ref{fig:duty} for $\Delta=0$ (other detunings only decreased the force). We observe that the forcing term that does work is not drastically increased with pulsed (short duty cycle) pumping.

Note that this is assuming that only the zeroth order correction for $J_{n}(\beta)$ is necessary. It is possible that higher order corrections, like that used in the derivation from \cite{Painter_Singlebeam}, may be needed. We observe that the maximum average force resulting from a square wave input is lower compared with the sinusoidal input of the same average power, which is expected as the power of the square input is spread into more Fourier components.

\end{document}